\documentclass{iopjournal}
\usepackage{amsmath, amssymb}
\usepackage{braket}
\usepackage{array}
\usepackage{xcolor}
\newcommand{\rev}[1]{\textcolor{blue}{#1}}

\begin{document}

\articletype{Article type} 

\title{ Anomalous superradiant phases in a quantum Rabi zigzag chain under a synthetic gauge field}

\author{Jun-Bo Tang, Lin-Jun Li, Bin Wei$^{*}$, and Yu-Yu Zhang$^{*}$}

\affil{Department of Physics, and Chongqing Key Laboratory for strongly coupled Physics,  Chongqing University, Chongqing 401330, China}

\affil{$^*$Author to whom any correspondence should be addressed.}

\email{weibin@cqu.edu.cn; yuyuzh@cqu.edu.cn}


\begin{abstract}
We present the phase diagram of a one-dimensional quantum Rabi zigzag chain under an artificial gauge field, which is implemented via phase-modulated next-nearest-neighbor (NNN) hopping. For a system size of $N=8$, we analytically determine the order parameters and critical thresholds, identifying five distinct superradiant phases. For strong nearest-neighbor (NN) coupling, the system exhibits either a ferro-superradiant (FSR) or an antiferro-superradiant (AFSR) phase depending on its sign. Increasing NNN coupling induces a stripe antiferro-superradiant (sAFSR) phase with antiparallel NNN and parallel NN displacements, which is distinct from the conventional AFSR phase. Tuning the magnetic flux drives the system into chiral or  antichiral superradiant phases with opposite ground-state photon currents, which are different from the conventional superradiant phases. Remarkably, geometric frustration within each triangular plaquette stabilizes chiral phases even at small magnetic flux while suppressing the sAFSR phase, which vanishes for $N=6$. Mapping the system onto an effective magnetic Hamiltonian yields an analogous phase diagram featuring distinct magnetic phases, offering insight into the anomalous superradiant phases. This equivalence establishes a quantum optical platform for simulating chiral magnetism.

\end{abstract}

\section{Introduction}
Light-matter interactions realized in cavity and circuit QED have given rise to a new class of quantum many-body systems~\cite{NaturePhysics2006,PhysRevLett2007,PhysRevLett2012}, where high tunability and control render these platforms ideal for investigating quantum phases of matter~\cite{PhysRevLett2014DK,PhysRevLett2014,PhysRevLett2011dk}. Considerable efforts have consequently been devoted to exploring quantum phase transitions within these synthetic quantum systems. A superradiant phase transition (SPT)~\cite{PhysRev.93.99,Emary03,PhysRevA2008} is a hallmark of collective light–matter interactions, manifesting as a cooperative burst of photons emitted by an ensemble of atoms, and has been realized in both Bose-Einstein condensates~\cite{nature2010} and Fermi gases~\cite{science2021,PhysRevLett.112.143004}.  Remarkably, such an SPT has also been identified in the quantum Rabi model for a single atom in proper limits~\cite{Ashhab2013,hwang2015,liu2017,chen2020,NCcai2021,chen2021,PhysRevA.111.043716}, providing a promising avenue for studying exotic phases and critical behavior in finite-component systems.

External fields provide a powerful route to engineer and utilize emerging phases. Recent experimental and theoretical efforts have demonstrated the synthetic  magnetic fields for neutral ultracold atoms~\cite{PhysRevLett.97.240401,bloch2012,dalibard2011,fu2014} and photonic
systems~\cite{roushan2017,wang2016,umucalilar2012}. In this context, interacting photon systems have emerged as an innovative
 platform for deepening the understanding of 
quantum many-body phases in the presence of artificial
magnetic fields. Evidence for this is provided by the emergence
of exotic phases of matter, such as chiral ground-state
currents of photons in a $3$-qubit loop~\cite{roushan2017}, the fractional quantum Hall physics in a Jaynes-Cummings-Hubbard lattice~\cite{hayward2012} and chiral phases in a quantum Rabi triangle~\cite{PhysRevLettzhang2021}. Besides artificial magnetic fields, considerable efforts have also been devoted to engineering exotic superradiant phases driven by additional geometric frustration in light-atom coupling settings, exemplified by Meissner-like superradiant phases with photon vortices~\cite{qj5x-t71k}, frustrated superradiant phases~\cite{PhysRevLett.128.163601, PhysRevA.110.013713}, and their quantum-magnetism analogues~\cite{PhysRevLettzhang2022}.

We investigate a one-dimensional quantum Rabi zigzag chain, where each cavity realizes a quantum Rabi model and photons hop between nearest-neighbor (NN) and next-nearest-neighbor (NNN) cavities. A synthetic gauge field is realized in each triangular plaquette by modulating the NNN hopping with a phase in circuit QED. In the presence of the magnetic flux, the quantum Rabi zigzag chain undergoes a second-order phase transition from the normal phase to a chiral superradiant (CSR) or antichiral superradiant (ACSR) phase depending on the sign of the NN hopping strength, which exhibits opposite ground-state photon currents. For positive NN hopping, we identify an antiferro-superradiant (AFSR) phase with antiparallel displacements along the NN cavity direction, while a stripe antiferro-superradiant (sAFSR) phase emerges with antiparallel displacements along the NNN direction for strong NNN coupling. The sAFSR is suppressed in favor of a chiral phase in the $N=6$ system with frustrated  sublattices. This reveals that geometric frustration in each triangular plaquette stabilizes the chiral phases even for weak flux.

\section{A synthetic gauge field in a quantum Rabi zigzag chain }
\begin{figure}[ht]
	\centering
\includegraphics[scale=0.25]{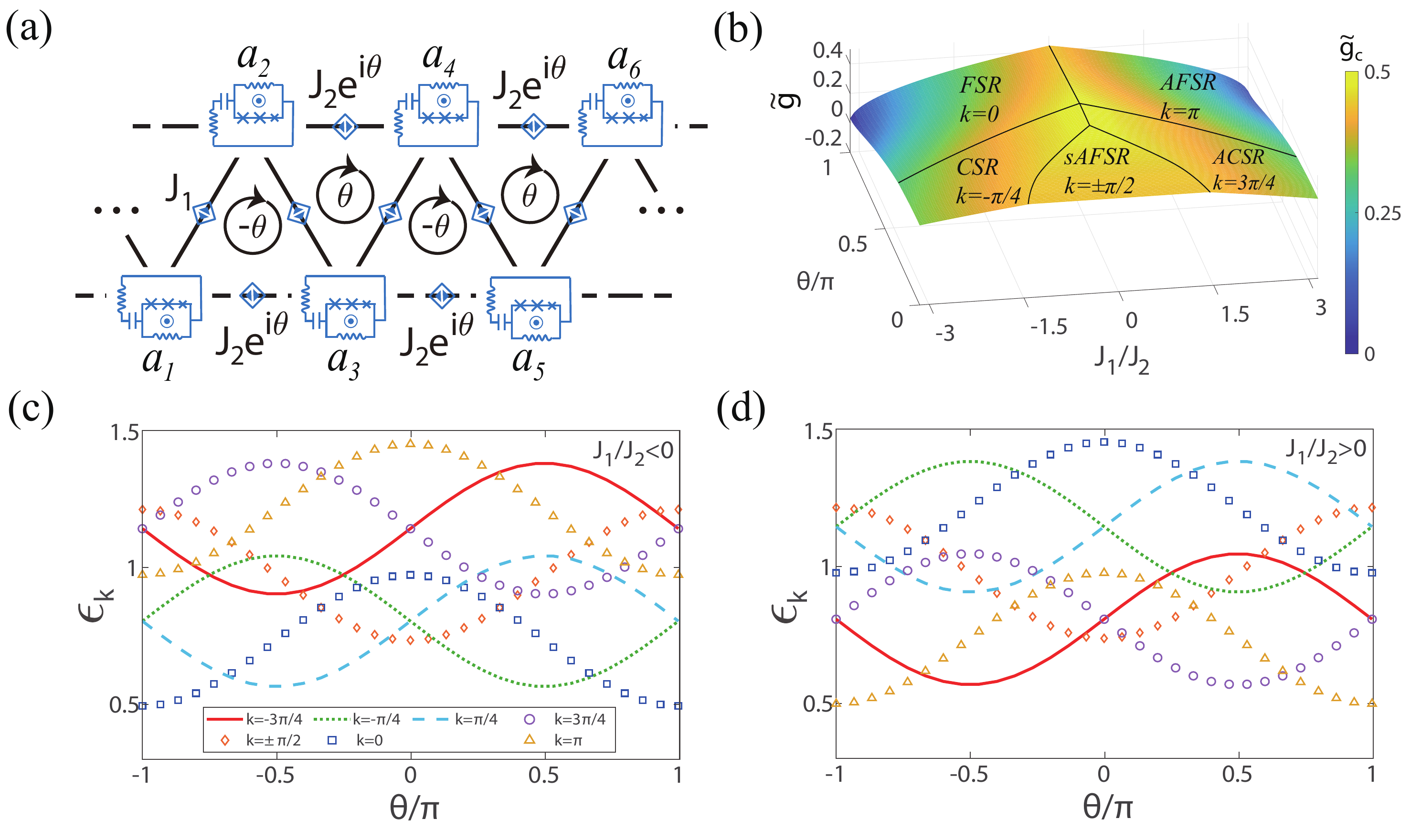}
\caption{(a) Schematic diagram of a quantum Rabi chain system realized in a circuit QED. Each LC resonator is strongly coupled to a qubit with a capacitor and is described by the quantum Rabi Hamiltonian. An artificial gauge field can be realized by a periodic modulation of the photon hopping between NNN resonators. (b) Ground-state phase diagram in the $\tilde{g}_{c}-\theta-J_1/J_2$ plane for system size $N=8$. The critical surface $\tilde{g}_{c}(k)$ analytically captures the boundary between the NP and superradiant phases: FSR, AFSR, sAFSR, CSR and ACSR phases. The solid lines indicate the first-order phase boundaries separating distinct superradiant phases. The excitation energy $\varepsilon_k$ as a function of $\theta$ at $\tilde{g} = 0.1$ for $J_1/J_2=-1$ (c) and for $J_1/J_2=1$ (d). In all our calculations, we set $\eta=\Delta/\omega=50$ and $J_2 = 0.12$, as well as choosing $\omega = 1$ as the unit for frequency.} \label{phase}
\end{figure}

We propose a one-dimensional quantum Rabi zigzag chain implemented in a circuit-QED architecture in Fig.~\ref{phase}. The chain consists of coupled resonators arranged in a zigzag configuration, where photons hopping between NN and NNN resonators can be realized by capacitive coupling. Each LC resonator, strongly coupled to a qubit with a capacitor, is described by the quantum Rabi Hamiltonian
\begin{equation}
 H_{R,n} = \omega_n a_n^{\dagger} a_n +g_n(t)(a_n^{\dagger} + a_n) \sigma_n^x + \frac{\Delta}{2} \sigma_n^z,   
\end{equation}
where $a_n^{\dagger} (a_n)$  is the bosonic creation (annihilation) operator of the $n$th single-mode cavity with the frequency $\omega_n$, and $\sigma_n^k$ ($k = x$, $y$, and $z$ are the Pauli matrices at site $n$ with energy splitting $\Delta$, and the atom-cavity coupling strength is modulated as $g_n(t)=g
\cos[(\omega_n-\omega)t]$. In circuit QED, NN cavities are capacitively coupling with tunable strength $J_{n,n+1}(t)$, whereas NNN cavities are inductively coupled through a mutual coupler loop with strength $J_{n,n+2}(t)$. The Hamiltonian of the quantum Rabi zigzag chain is given by 
\begin{eqnarray}
H_{RZ}=\sum_{n=1}^{N}H_{R,n} +J_{n,n+1}(t)(a_n^{\dagger}a_{n+1}+a_na_{n+1}^{\dagger})+J_{n,n+2}(t)(a_n^{\dagger}a_{n+2}+a_na_{n+2}^{\dagger}).
\end{eqnarray}

An artificial magnetic field is generated by periodically modulating the photon hopping terms~\cite{roushan2017,PhysRevLettzhang2021}, in analogy with the magnetic field acting on electrons. We consider time-dependent hopping strengths $J_{n,n+1}(t)=J_1\cos[(\omega_{n+1}-\omega_{n})t]$ and $J_{n,n+2}(t)=J_2\cos[(\omega_{n+2}-\omega_{n})t+\theta]$. In the rotating frame via the unitary transformation $U=e^{itH'_0}$ with $H'_0=\sum_{n=1}^{N}(\omega_n-\omega) a_n^{\dagger}a_n$, the NNN photon hopping terms, $H_{\texttt{NNN}}=UJ_{n,n+2}(t)(a_n^{\dagger}a_{n+2}+a_na_{n+2}^{\dagger})U^{\dagger}$, is given by 
\begin{eqnarray}
H_{\texttt{NNN}}=J_2\cos[(\omega_{n+2}-\omega_{n})t+\theta][e^{-i(\omega_{n}-\omega_{n+2})t}a_{n}^{\dagger}a_{n+2}+h.c.]\approx J_2(e^{i\theta}a_{n}^{\dagger}a_{n+2}+e^{-i\theta}a_na_{n+2}^{\dagger}),
\end{eqnarray}
where the fast oscillating terms $e^{2i(\omega_{n+2}-\omega_{n})+\theta}$ are neglected. Similarly, the effective NN hopping Hamiltonian $H_{\texttt{NN}}=UJ_{n,n+1}(t)(a_n^{\dagger}a_{n+1}+a_na_{n+1}^{\dagger})U^{\dagger}$ becomes
\begin{eqnarray}
H_{\texttt{NN}}=J_1\cos[(\omega_{n+1}-\omega_{n})t][e^{-i(\omega_{n}-\omega_{n+1})t}a_{n}^{\dagger}a_{n+1}+a_na_{n+1}^{\dagger}.]\approx J_1(a_{n}^{\dagger}a_{n+1}+a_na_{n+1}^{\dagger}).
\end{eqnarray}

We obtain the effective time-independent Hamiltonian by $H_{RZ}=UHU^{\dagger}-iU\frac{d}{dt}U^{\dagger}$, giving
\begin{eqnarray}\label{ham1}
H_{RZ}&=&\sum_{n=1}^{N}H_{R,n}+J_1(a_{n}^{\dagger}a_{n+1}+a_na_{n+1}^{\dagger})
+J_2(e^{i\theta}a_{n}^{\dagger}a_{n+2}+e^{-i\theta}a_na_{n+2}^{\dagger}),
\end{eqnarray}
where the quantum Rabi Hamiltonian of each resonator coupled to a qubit is
\begin{equation}
 H_{R,n}=\omega a_{n}^{\dagger}a_n+g(a_{n}^{\dagger}+a_n)\sigma_n^x+\frac{\Delta}{2}\sigma_n^z.   
\end{equation} 
The quantum Rabi model has been proven to undergo a superradiant phase transition in the frequency limit $\eta=\Delta/\omega \to \infty$~\cite{Ashhab2013,hwang2015,chen2020}, which plays the role of a thermodynamic limit that the spectral gap closes at the critical value. The scaled coupling strength is given by $\tilde{g}=g/\sqrt{\omega \Delta}$.

The photons hopping amplitude between NNN cavities in Eq.(\ref{ham1}) is complex with the phase $\theta$. This phase can be interpreted as arising from an artificial
vector potential $A(r)$, such that $\theta =\int_{\mathbf{r}_{n}}^{\mathbf{r}_{m}}$ $A(\mathbf{r})d\mathbf{r}$~\cite{hayward2012}. It yields a gauge-invariant effective magnetic flux $\phi=\theta$ in each upward and downward triangular plaquette, under periodic boundary condition $a_{N+1} = a_1$.

 \begin{figure}[ht]
	\centering
\includegraphics[scale=0.25]{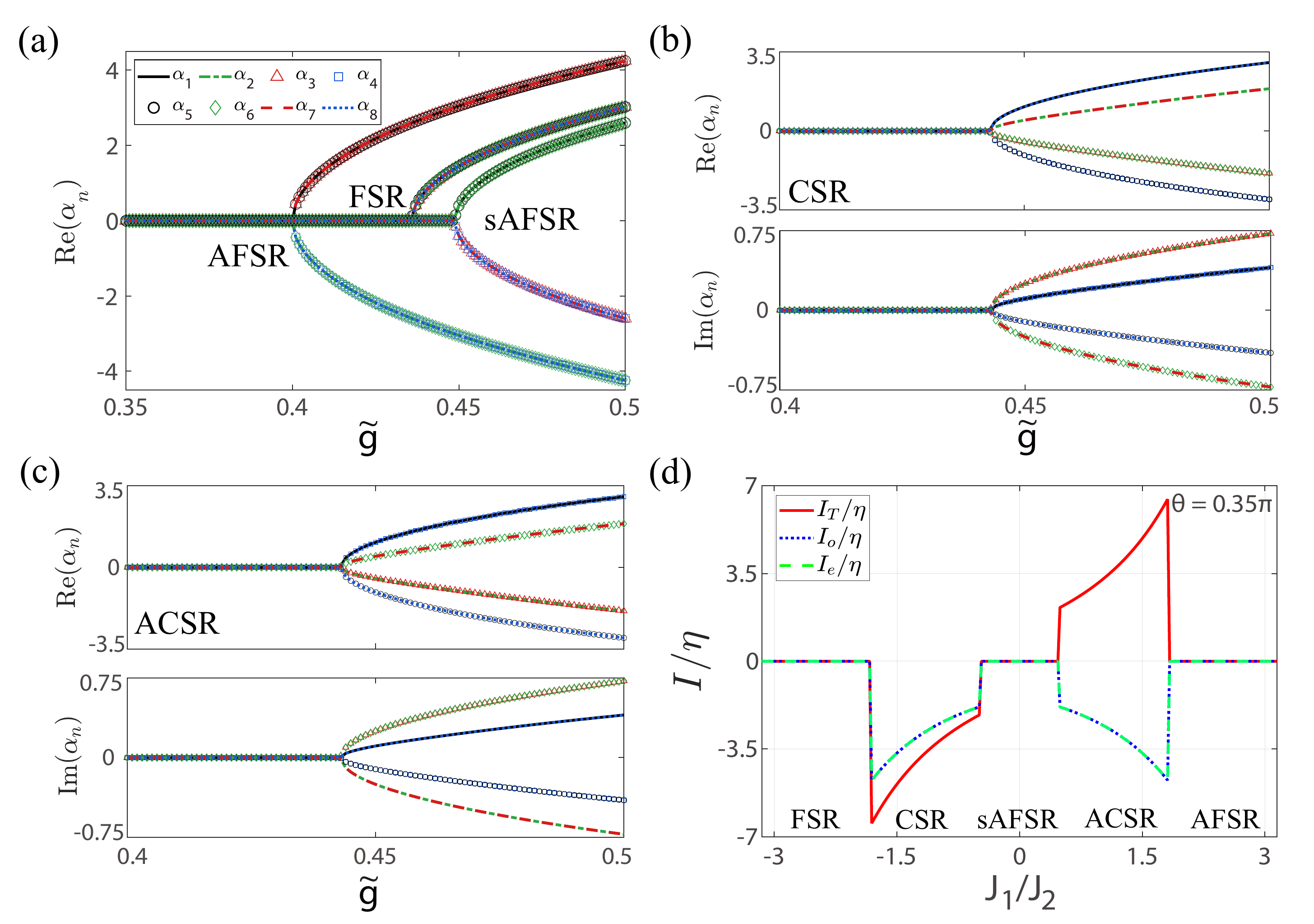}
\caption{(a) Order parameter $\alpha_n$ as a function of the scaled coupling strength 
$\widetilde{g}$ for $N=8$ in the FSR ($J_1/J_2=-1,\theta=0.5\pi$), AFSR ($J_1/J_2=1.5,\theta=0.5\pi$), and sAFSR ($J_1/J_2=-1,\theta=0.2\pi$) phases, respectively. Real and imaginary parts of $\alpha_n$ in the CSR ($J_1/J_2 = -1$) (b) and ACSR ($J_1/J_2 = 1$) (c) phases at $\theta= 0.3\pi$. (d) Scaled photon current along the chain $I_T/\eta$, along with sublattice currents $I_{o(e)}/\eta$ versus $J_1/J_2$ across the superradiant phase transitions with $\theta=0.35\pi$ and $\tilde{g}=0.7$.} \label{alpha}  
\end{figure}

\section{Normal phase}\label{normal phase}
In the weak-coupling regime, the number of excitations in the ground state tends to zero, and the system is in the normal phase. To determine the low-energy excitation energies, we
perform a Schrieffer-Wolff transformation using the unitary operator $U = \prod_{n=1}^{N} \exp\left[ -i\sigma_n^y \tilde{g} \sqrt{\omega/\Delta} (a_n^\dagger + a_n) \right]$. Under this transformation, the Hamiltonian in Eq.(\ref{ham1}) becomes
\begin{eqnarray}
H_{\mathrm{NP}}=\sum_{n=1}^N\omega a_n^\dagger a_n +\frac{\Delta}{2}\sigma_n^z +\omega\widetilde{g}^{\,2}(a_n^\dagger+a_n)^2\sigma_n^z+J_1 \left( a_n^\dagger a_{n+1}+a_{n+1}^\dagger a_{n} \right)+J_2( e^{i\theta} a_n^\dagger a_{n+2} +H.c.),
\end{eqnarray}
where higher-order terms are neglected in the limit $\Delta/\omega\to \infty$. Since the transverse term involving $\sigma_n^x$ is eliminated, the two atomic levels are decoupled. Thus, the low-energy effective
Hamiltonian can be obtained by projecting onto the subspace
of the lower atomic level $|\downarrow\rangle$, i.e.\rev{,} $H_{\mathrm{NP}}^{\downarrow}=\langle\downarrow| H_{\mathrm{NP}}|\downarrow\rangle $. This yields
\begin{eqnarray}\label{hamnp}
H_{\mathrm{NP}}^{\downarrow} 
= \sum_{n=1}^{N} \omega a_n^\dagger a_n - \omega \tilde{g}^2 (a_n^\dagger + a_n)^2 
+ J_1 \left( a_n^\dagger a_{n+1}+a_{n+1}^\dagger a_{n} \right)+J_2 \left( e^{i\theta} a_n^\dagger a_{n+2} +e^{-i\theta} a_{n+2}^\dagger a_{n} \right)-\frac{\Delta}{2}.
\end{eqnarray}

Applying the Fourier transformation $a_n = \sum_k e^{ink} a_k / \sqrt{N}$, we obtain
\begin{equation}\label{hamq}
 H_{\mathrm{NP}}^{\downarrow} = \sum_k\omega_k a_k^\dagger a_k - \omega \tilde{g}^2 (a_k^\dagger a_{-k}^\dagger + a_k a_{-k})+E_0, 
\end{equation}
where the frequencies are $\omega_k = \omega - 2\omega \tilde{g}^2 + 2[J_1 \cos k + J_2 \cos(\theta+2k)]$ with the quasimomentum $k = 0, 2\pi/N, ..., 2(N-1)\pi/N$, and $E_0 = \sum_{n=1}^{N}\left[ -\Delta/2 - \omega \tilde{g}^2 + (\omega + J_1 + J_2)\omega\tilde{g}^2/\Delta \right]$. We apply the unitary squeezing transformation \(S =  \prod_{k}\exp\left[ \lambda_k (a_k^\dagger a_{-k}^\dagger - a_k a_{-k}) \right]\) with the variational parameter $\lambda_k$. Under this transformation, the Hamiltonian in Eq.(\ref{hamq}) becomes 
	\begin{align}
		H_{\mathrm{NP}}^{\downarrow}
&= \sum_k\left[\omega_k\cosh^2(2\lambda_k)-2\omega\tilde{g}^2\sinh(2\lambda_k)\cosh(2\lambda_k)\right]a_k^\dagger a_k\nonumber\\
&\quad+\left[\omega_k\sinh^2(2\lambda_k)-2\omega\tilde{g}^2\sinh(2\lambda_k)\cosh(2\lambda_k)\right]a_{-k}^\dagger a_{-k}+\frac{1}{2}V_{\mathrm{off}}+E_g ,
	\end{align}
	where the off-diagonal term is given by $V_{\mathrm{off}}= \sum_k \left[ (\omega_k + \omega_{-k}) \sinh(2\lambda_k) \cosh(2\lambda_k) - 2\omega \tilde{g}^2 (\cosh^2 2\lambda_k + \sinh^2 2\lambda_k) \right](a_k a_{-k} + a_k^\dagger a_{-k}^\dagger)$. The squeezing  parameter is determined by imposing $V_{\mathrm{off}} = 0$, yielding $\lambda_k =1/8\ln[(\omega_k + \omega_{-k} + 4\omega \tilde{g}^2)/(\omega_k + \omega_{-k} - 4\omega \tilde{g}^2])$.
	The diagonal Hamiltonian takes the form $H_{\mathrm{NP}}^{\downarrow} = \sum_k \varepsilon_k a_k^\dagger a_k + E_g$, where the ground-state energy is $ E_g=\sum_k (\varepsilon_k-\omega_k)/2+E_0$. The excitation energy is analytically given by the dispersion relation
	\begin{equation}
		\varepsilon_k = \frac{1}{2} \left[ \sqrt{ (\omega_k + \omega_{-k})^2 - 16\omega^2 \tilde{g}^4 } + \omega_k - \omega_{-k} \right].
	\end{equation}
    As the coupling $\tilde{g}$ increases, each branch $\varepsilon_k$ decreases monotonically and vanishes at the critical value 
	\begin{equation}
		\tilde{g}_{c}(k) = \sqrt{ \frac{ (\omega + 2J_1 \cos k + 2J_2 \cos\theta \cos 2k)^2 -(2J_2 \sin\theta \sin 2k)^2 }{ 4\omega(\omega + 2J_1 \cos k + 2J_2 \cos\theta \cos 2k) } }.
	\end{equation}
    The critical coupling implies $k$-dependent phase boundaries. 
    
    Fig.~\ref{phase} shows the phase boundaries and excitation energies $\varepsilon_k$ for $N=8$. These distinct momentum-dependent critical couplings $\tilde{g}_{c}(k)$ signal different quantum phases for $ \tilde{g}>\tilde{g}_{c}(k)$, as illustrated in Fig.~\ref{phase}(b). For $J_1 /J_2<0$, the lowest-energy mode occurs at momenta $k=0$, $\pm\pi/4$, $\pm\pi/2$, depending on the artificial flux $\theta$, as shown in Fig.~\ref{phase}(c). The critical surfaces $\tilde{g}_{c}(k=0)$ and $\tilde{g}_{c}(k=-\pi/4)$ merge along a critical line, yielding the critical flux value 
    \begin{equation}
    \theta_{c,1}^- = \cos^{-1}\left[ \frac{ \omega+\sqrt{2}J_1 - \sqrt{ \left(\omega+\sqrt{2}J_1\right) \left[\omega+(8-3\sqrt{2})J_1\right]+16J_2^2}}{4J_2}\right].
    \end{equation}
  Additionally, the other critical flux $\theta_{c,2}^{-}$ and the associated critical hopping $J_{1,c}^{-}$ can be obtained from $\tilde{g}_{c}(k=\pm\pi/2)=\tilde{g}_{c}(k=-\pi/4)$, which yields
   \begin{align}
\theta_{c,2}^-& = \cos^{-1}\left[ \frac{ -(\omega+\sqrt{2}J_1)+ \sqrt{ \omega^2-2\sqrt{2}\omega J_1 - 6 J_1^2 +16 J_2^2}}{4J_2}\right],\\
J_{1,c}^- &=\frac{ -(\omega+2J_2\cos\theta)+ \sqrt{ \omega^2-4\omega J_2\cos\theta - 12\omega J_2^2\cos^2\theta +16 J_2^2}}{2\sqrt{2}}.
   \end{align}
  
  For $J_1 /J_2>0$, by contrast, the lowest-energy mode occurs at  $k=\pi$, $\pm3\pi/4$, and $\pm\pi/2$. The boundary between the critical surfaces $\tilde{g}_{c}(k=\pi)$ and $\tilde{g}_{c}(k=3\pi/4)$ is determined by
    \begin{align}
\theta_{c,1}^+ =\cos^{-1}\left[\frac{\omega-\sqrt{2}J_1-\sqrt{\left(\omega-\sqrt{2}J_1\right) \left[\omega-(8-3\sqrt{2})J_1\right] +16J_2^2 } }{ 4J_2 } \right].
    \end{align}
   Additionally, equating $\tilde{g}_{c}(k=\pm\pi/2)=\tilde{g}_{c}(k=3\pi/4)$ gives another critical flux 
  \begin{align}
\theta_{c,2}^+ &=\cos^{-1}\left[\frac{-(\omega-\sqrt{2}J_1)+\sqrt{\omega^2+2\sqrt{2}\omega J_1-6 J_1^2 +16 J_2^2} }{ 4J_2 } \right]\\
 \end{align}
and the critical hopping  
 \begin{align}
J_{1,c}^+& = \frac{ (\omega+2J_2\cos\theta)-\sqrt{ \omega^2-4\omega J_2\cos\theta - 12\omega J_2^2\cos^2\theta +16 J_2^2}}{2\sqrt{2}}.
  \end{align}

\section{Superradiant phases}

For $\tilde{g}>\tilde{g}_{c}(k)$, the system enters the superradiant regime, where the photon population in each cavity becomes macroscopic. The bosonic mode is shifted as $a_n\to \widetilde{a}_n + \alpha_n$ with a complex displacement $\alpha_n = A_n + iB_n$. The Hamiltonian can be written as $H_{\mathrm{SR}} = H_s + H_1$, where
	\begin{align}
		H_s &= \sum_{n=1} ^Ng (\alpha_n^* + \alpha_n) \sigma_n^x + \frac{\Delta}{2} \sigma_n^z, \label{eq:22} \\
		H_1 &= \sum_{n=1}^N\omega (\widetilde{a}_n^\dagger + \alpha_n^*) (\widetilde{a}_n + \alpha_n) +g (\widetilde{a}_n^\dagger + \widetilde{a}_n) \sigma_n^x
		+ J_1  (\widetilde{a}_n^\dagger + \alpha_n^*) \left[ (\widetilde{a}_{n+1} + \alpha_{n+1}) + (\widetilde{a}_{n-1} + \alpha_{n-1}) \right] \nonumber \\
		&+ J_2  (\widetilde{a}_n^\dagger + \alpha_n^*) \left[ e^{i\theta} (\widetilde{a}_{n+2} + \alpha_{n+2}) + e^{-i\theta} (\widetilde{a}_{n-2} + \alpha_{n-2}) \right].
	\end{align}
	The atomic part $H_s$ can be diagonalized in terms of the eigenstates $\rev{|}+z\rangle = \cos\gamma_n|e\rangle + \sin\gamma_n|g\rangle$ and $\rev{|}-z\rangle = -\sin\gamma_n|e\rangle + \cos\gamma_n|g\rangle$ with $\tan 2\gamma_n = 4g A_n /\Delta$. In the basis of $\{\ket{+z}, \ket{-z}\}$, the Hamiltonian $H_\mathrm{SR}$ can be expressed as
	\begin{align}\label{SR1}
		H_{\mathrm{SR}} &= \sum_{n=1}^N\omega \widetilde{a}_n^\dagger \widetilde{a}_n + \frac{\Delta_n'}{2} \tau_n^z + g_n' (\widetilde{a}_n^\dagger + \widetilde{a}_n) \tau_n^x +J_1 \widetilde{a}_n^\dagger (\widetilde{a}_{n+1} + \widetilde{a}_{n-1}) \nonumber\\
        &+J_2 \widetilde{a}_n^\dagger ( e^{i\theta} \widetilde{a}_{n+2} + e^{-i\theta} \widetilde{a}_{n-2})+ V_{\mathrm{off}} + E_0, 
	\end{align}
	where the renormalized coupling strength is $g_n' = g \Delta /\Delta_n'$, the renormalized atomic frequency is $\Delta_n' = \sqrt{\Delta^2 + 16g^2 A_n^2}$, and the transformed Pauli matrix is $\tau_n^z = \sigma_n^z\Delta/\Delta_n'  + 4g \sigma_n^xA_n/\Delta_n'$. The constant energy term is $E_0= \sum_{n=1} ^N\omega |\alpha_n|^2 + J_1 \sum_{n=1} ^N\alpha_n^* (\alpha_{n+1} + \alpha_{n-1})+ J_2 \sum_{n=1}^{N} \alpha_n^* (e^{i\theta} \alpha_{n+2} + e^{-i\theta} \alpha_{n-2})$, and the off-diagonal term is given by
	\begin{align}
		V_{\mathrm{off}} &= \sum_{n=1}^N \left[ \omega (\widetilde{a}_n^\dagger \alpha_n + \alpha_n^* \widetilde{a}_n) + g (\widetilde{a}_n^\dagger + \widetilde{a}_n) \sin(2\gamma_n) \tau_n^z \right] + J_1 \left[ \widetilde{a}_n^\dagger (\alpha_{n+1} + \alpha_{n-1}) + h.c. \right] \nonumber\\
        &+J_2 \left[ \widetilde{a}_n^\dagger (e^{i\theta} \alpha_{n+2} + e^{-i\theta} \alpha_{n-2}) + h.c. \right]. 
	\end{align}
    By requiring the off-diagonal term $V_{\mathrm{off}}$ to vanish, we obtain the effective Hamiltonian $H_{\mathrm{SR}}$ (\ref{SR1}) in the superradiant regime, which has the same structure as Hamiltonian (\ref{ham1}). Thus, by employing the same procedure used to derive $H_{\mathrm{NP}}^{\downarrow}$, we obtain the effective low-energy Hamiltonian in the superradiant phases by projecting onto the lower atomic subspace
	\begin{align}
		H_{\mathrm{SR}}^{\downarrow} = \sum_{n=1}^N\left[ \omega \widetilde{a}_n^\dagger \widetilde{a}_n - \frac{{g_n'}^2}{\Delta_n'} (\widetilde{a}_n^\dagger + \widetilde{a}_n)^2 \right] +J_1 \widetilde{a}_n^\dagger (\widetilde{a}_{n+1} + \widetilde{a}_{n-1})+J_2 \widetilde{a}_n^\dagger ( e^{i\theta} \widetilde{a}_{n+2} + e^{-i\theta} \widetilde{a}_{n-2}) + E_g\rev{,}
	\end{align}
    where the ground-state energy is $E_g=E_0-\sum_{n=1}^N\Delta_n'/2$. Diagonalizing the above quadratic Hamiltonian, we obtain the excitation spectrum in the superadiant phases.

   The displacement $\alpha_n$ can be determined by requiring the off-diagonal term $V_{\mathrm{off}}=0$, which yields $\omega \alpha_n - g \sin(2\gamma_n) + J_1 (\alpha_{n+1} + \alpha_{n-1}) + J_2 (e^{i\theta} \alpha_{n+2} + e^{-i\theta} \alpha_{n-2}) = 0$. The real and imaginary parts of $\alpha_n$ then satisfy the following equations
        \begin{align}\label{An}
            ( \omega-4g^2/\Delta_n') A_n + J_1 (A_{n+1} + A_{n-1})+ J_2 \cos\theta (A_{n+2} + A_{n-2}) + J_2 \sin\theta ( B_{n-2}-B_{n+2}) = 0, 
        \end{align}
        \begin{align}\label{Bn}
            \omega B_n + J_1 (B_{n+1} + B_{n-1})+ J_2 \sin\theta (A_{n+2} - A_{n-2}) + J_2 \cos\theta (B_{n+2} + B_{n-2}) = 0.
        \end{align}
The order parameter $\alpha_n$ obtained from Eqs. (\ref{An})-(\ref{Bn}) characterizes distinct superradiant phase\rev{s} in the $N=8$ system.

       \begin{figure}[ht]
		\centering
		\includegraphics[scale=0.25]{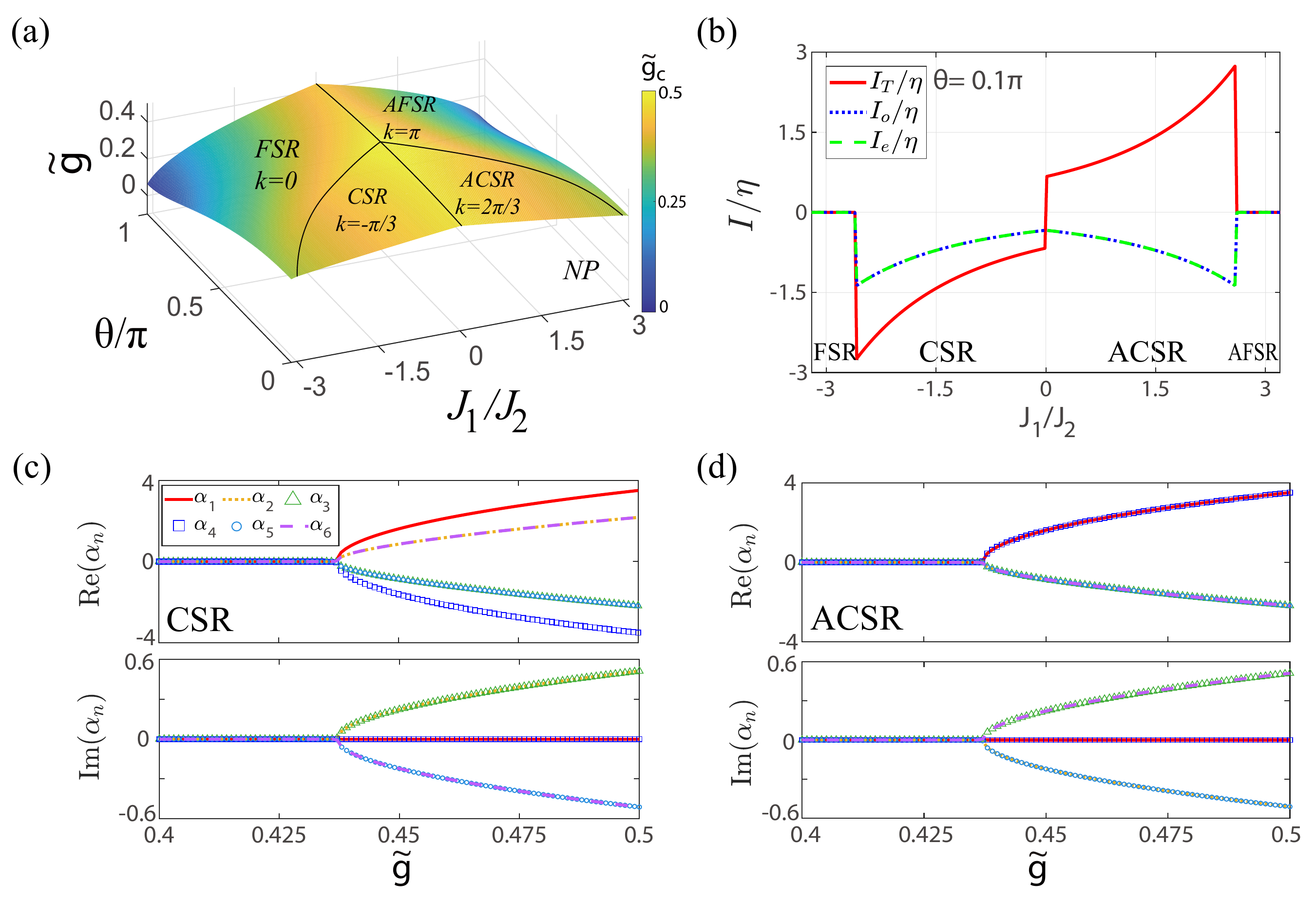}
		\caption{(a) Phase diagram in the $\tilde{g}_{c}-\theta-J_1/J_2$ plane for the quantum Rabi zigzag chain for $N=6$. Critical surface $\tilde{g}_{c}$ separates the NP and superradiant phases (FSR, AFSR, CSR and ACSR phases). (b) Scaled photon current along the chain $I_T/\eta$ and sublattice currents $I_{o}/\eta$ and $I_{e}/\eta$ as a function of $J_1/J_2$ with $\theta=0.1\pi$ and $\tilde{g}=0.7$. Real and imaginary parts of $\alpha_n$ as a function of $\tilde{g}$ for the CSR ($J_1/J_2=-1$) (c) and ACSR ($J_1/J_2=1$) (d) phases at $\theta=0.2\pi$, respectively. }\label{phase6}
	\end{figure}

(\textit{i}) \textit{Ferro-superradiant (FSR) phase. } For $|\theta_{c,1}^{-}| \leq |\theta| \leq \pi $ and $ J_1/J_2 < 0$, the order parameter $\alpha$ is real and explicitly given by $\alpha_n= \pm\sqrt{16g^4/(\omega + 2J_1 + 2J_2 \cos\theta)^2 - \Delta^2 }/(4g)$. Thus, $\alpha$ is the same for all cavities in Fig. \ref{alpha}(a). The ground state is doubly degenerate as a consequence of $Z_2$ symmetry breaking. This configuration is analogous to the ferromagnetic ordering in spin systems and is therefore referred to as the FSR phase. $\alpha_n$ grows from zero, indicating a
second-order phase transition from the NP to the FSR phase.

(\textit{ii}) \textit{Antiferro-superradiant (AFSR) phase.} For $|\theta_{c,1}^{+}| \leq |\theta| \leq \pi $ and $ J_1/J_2 > 0$, the order parameter $\alpha$ is also real but acquires opposite signs  between NN cavities, giving $\alpha_n=-\alpha_{n\pm1}$. The displacement is analytically obtained as $\alpha_n =\pm\sqrt{ 16g^4/(\omega - 2J_1 + 2J_2 \cos\theta)^2 - \Delta^2 }/(4g)$. Fig. \ref{alpha}(a) shows that the NN cavities have the opposite signs of $\alpha_n$. This configuration is analogous to antiferromagnetic ordering in spin systems and is therefore referred to as the AFSR phase.

(\textit{iii}) \textit{Chiral superradiant (CSR) phase.}  For $|\theta_{c,2}^{-}| \leq |\theta| \leq |\theta_{c,1}^{-}|$ and $J_1<J_{1,c}^-<0$, the order parameter $\alpha_n$ is complex and depends on $n$. Fig. \ref{alpha}(b) shows one of the solutions of $\alpha_n$, for which the real and imaginary parts satisfy $A_1=A_8=-A_4=-A_5$, $A_2 = -A_3 = -A_6 = A_7$, $B_1=B_4=-B_5=-B_8$, and $B_2= B_3 = -B_6 = -B_7$. Due to the $C_4$ symmetry of the even and odd sublattices and the underlying $Z_2$ symmetry, the ground state exhibits eightfold degeneracy. The real and imaginary parts of $\alpha_n$ are analogous to the in-plane spin components in the $xy$ plane. This phase is referred to as the CSR phase. Fig. \ref{alpha}(b) show $\alpha_n$ grows with increasing coupling strength for $\tilde{g}>\tilde{g}_{c}$, indicating a second-order phase transition from NP to CSR phase. Remarkably, the CSR phase is induced by the artificial magnetic field, yet geometric frustration within each triangular plaquette plays a crucial role in stabilizing the chiral phase even for small $\theta$. 

(\textit{iv}) \textit{Antichiral superradiant (ACSR) phase.} For $|\theta_{c,2}^{+}| \leq |\theta| \leq |\theta_{c,1}^{+}| $ and $J_1>J_{1,c}^+>0$, the order parameter $\alpha_n$ is complex. In contrast to the CSR phase, $\alpha_n$ has different solutions in Fig. \ref{alpha}(c), exhibiting $A_1=-A_8=A_4=-A_5$, $A_2 = A_3 = -A_6 = -A_7$, $B_1=-B_4=-B_5=B_8$, and $B_2=- B_3 = -B_6 = B_7$. The positive NN hopping $J_1>0$ imposes antiferromagnetic coupling between the odd and even sublattices. As a result, for each $\alpha_n$ pair on the sublattices, such as sites ($1,2$), ($3,4$), ($5,6$), and ($7,8$), the real and imaginary parts acquire opposite orientations. This phase is referred to as the ACSR phase. 


(\textit{v}) \textit{Stripe antiferro-superradiant (sAFSR) phase.} For $0 \leq |\theta| \leq |\theta_{c,2}^{\pm}| $ and $J_{1,c}^-<J_1<J_{1,c}^+$, the NNN hopping $J_2$ dominates over the NN  coupling $J_1$. This results in opposite signs of $\alpha$ between NNN cavities in the even and odd sublattices, in contrast to the AFSR phase. It leads to $\alpha_n=-\alpha_{n\pm2}$, where $\alpha_n =\sqrt{16g^4 /(\omega-2J_2\cos\theta)^2 -\Delta^2 }/(4g)$. Fig. \ref{alpha}(a) shows the NNN cavities exhibit opposite $\alpha_n$. The system forms the sAFSR phase.

    We now examine the superradiant phases with geometric frustration in the $N=6$ system. Each sublattice consists of three cavities with $C_3$ symmetry, which gives rise to frustration in the sublattice geometry. Here, depending on the sign of the NN hopping $J_1$, the system exhibits either an FSR or an AFSR phase with a real $\alpha_n$ configuration. However, owing to the geometric frustration within the sublattices, an sAFSR phase with antiferromagnetic NNN ordering is suppressed. The system consequently favors a CSR or ACSR phase characterized by complex $\alpha_n$. Thus, unlike the $N=8$ case, the sAFSR phase is absent here. The corresponding ground-state phase diagram is shown in Fig.~\ref{phase6}(a), where the critical coupling value $\tilde{g}_{c}(k)$ depends on the momentum $k=0,\pm\pi/3,\pm2\pi/3$ and $\pi$. The order parameter $\alpha_n$ in the CSR and ACSR phases exhibits a specific pattern in which the imaginary parts satisfy $B_1=B_4=0$ while all others remain nonzero, as shown in Fig.~\ref{phase6}(c) and Fig.~\ref{phase6} (d). This structure arises from three-cavity frustration intrinsic to each sublattice, which is markedly different from that in the $N=8$ case.

     \section{Chiral current}
     We investigate photon currents in the zigzag cavity loops to characterize the chirality of the anomalous superradiant phases. Analogous to the classical continuity equation, the photon current operator along the chain is $\hat{I}_{T}=\sum_{n=1}^{N} i( a_{n}^{\dagger}a_{n+1}-a_{n+1}^{\dagger}a_{n})$, which is approximated as $I_{T}=\sum_{n=1}^{N} -2\text{Im}(\alpha_{n}^{*}\alpha_{n+1})$ in the mean-field approximation. Meanwhile, the photon currents in the even and odd sublattices are characterized by $I_{\nu}=\sum_{n=\nu}-2\text{Im}(\alpha_{n}^{*}\alpha_{n+2})$ with $\nu=(e,o)$. 

     Fig.~\ref{phase}(d) shows the scaled photon currents as functions of the hopping strength $J_1/J_2$ for $\theta=0.1\pi$. In the CSR phase, the currents in the even and odd sublattices $I_{e}$ and $I_o$ flow in the same direction as the total chain current $I_{T}$, whereas in the ACSR phase they flow in the opposite direction. By tuning the NN coupling from ferromagnetic ( $J_1<0$) to antiferromagnetic ( $J_1>0$), the total chain current $I_{T}$ in the ACSR phase reverses direction relative to that in the CSR phase. In contrast, the current jumps to zero in the FSR, AFSR and sAFSR phases, which is associated with the real $\alpha_n$. This indicates that the transitions between these superradiant phases are first order transitions. Similarly, the frustrated $N=6$ geometry also exhibits chiral currents in the CSR and ACSR phases, as shown in Fig.~\ref{phase6}(b).

\section{Effective magnetic model}\label{effective magnetic model}
We now map the strongly coupled light-atom interaction system onto an effective magnetic Hamiltonian. This mapping enables an interpretation of the phase diagram in the quantum optical model in terms of competing magnetic interactions. We apply the Holstein–Primakoff transformation, given by $S_n^z = S-a_n^\dagger a_n$ and $ S_n^- = a_n^\dagger \sqrt{2S - a_n^\dagger a_n}$. In the classical spin limit $S=\Delta/\omega \to \infty$, where $S_n^- \approx \sqrt{2S} a_n^\dagger$, the Hamiltonian in Eq.(\ref{hamnp}) maps onto the effective  magnetic Hamiltonian
	\begin{align}
		H_{\mathrm{M}} &= \sum_{n=1}^{N} \left[ -\omega  S_n^z-\frac{2\tilde{g}^2\omega}{S}(S_n^x)^2  \right] 
        + \frac{J_1}{S} (S_n^x S_{n+1}^x + S_n^y S_{n+1}^y) \nonumber\\&+ \frac{J_2}{S}\cos\theta (S_n^x S_{n+2}^x + S_n^y S_{n+2}^y) - \frac{J_2}{S} \sin\theta\sum_{n=1}^{N} \vec{z} \cdot (\vec{S}_n \times \vec{S}_{n+2}). 
	\end{align}
The first two terms describe the $z$-axis external field and the $x$-axis easy-axis anisotropy, respectively. The $J_1$ term represents the $XY$-Heisenberg interactions between NN cavities, which can be ferromagnetic or antiferromagnetic depending on the sign of $J_1$. Remarkably, the magnetic flux $\theta$ simultaneously gives rise to an $XY$-Heisenberg interaction of strength $J_2 \cos\theta$ and a Dzyaloshinskii-Moriya (DM) interaction of strength $J_2 \sin \theta$ between NNN cavities. In particular, the DM interaction favors noncollinear spin textures in chiral magnets~\cite{PhysRevLettzhang2022}, and reaches its maximum strength at $\theta=\pm\pi/2$. Unlike previous work~\cite{PhysRevLettzhang2022}, the DM and $XY$ couplings between NNN sites tuned by $\theta$ compete with the NN $XY$ interactions, leading to rich quantum phases.

\begin{figure}[ht]
		\centering
		\includegraphics[scale=0.3]{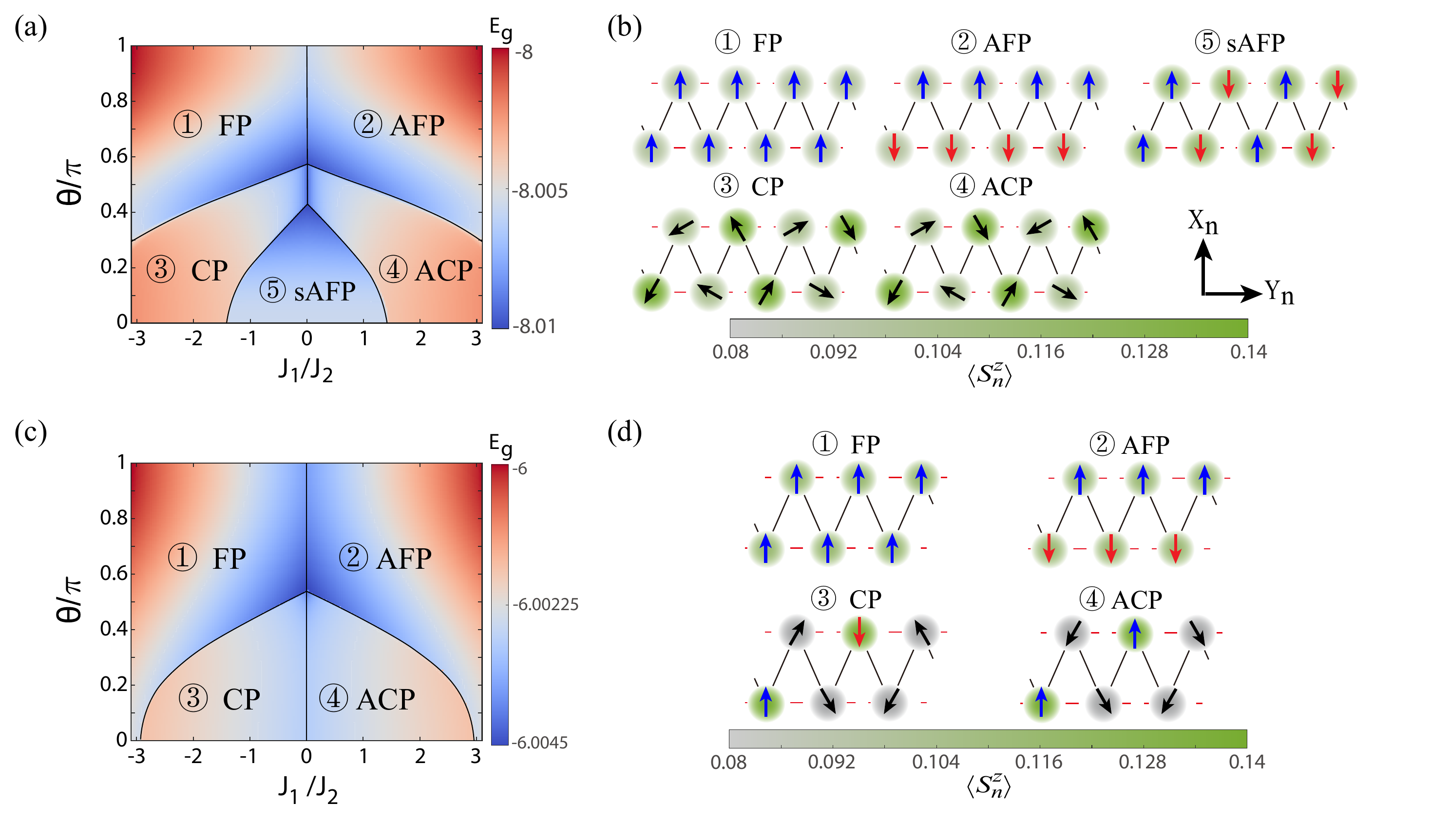}
		\caption{Phase diagram of the effective magnetic model in the $\theta-J_1/J_2$ plane for $N=8$ (a) and $N=6$ (c). The critical lines separate distinct magnetic phases. Corresponding magnetization configurations $(X_n,Y_n)$ are represented in the $xy$ plane for $N=8$ (b) and $N=6$ (d), where the color map indicates the expectation value $\langle S_n^z\rangle$ at each site.}\label{spin8}
	\end{figure}
The mean-field energy of the effective magnetic Hamiltonian is given by   
\begin{align}
		\frac{E}{\omega S} &= \sum_{n=1}^N \left[ -2\tilde{g}^2 X_n^2 - \sqrt{1 - X_n^2 - Y_n^2} \right] + \frac{J_1}{\omega} \sum_{n=1}^N (X_n X_{n+1} + Y_n Y_{n+1}) \nonumber \\
		&+ \frac{J_2 \cos\theta}{\omega} \sum_{n=1}^N (X_n X_{n+2} + Y_n Y_{n+2}) - \frac{J_2 \sin\theta}{\omega} \sum_{n=1}^N (X_n Y_{n+2} - Y_n X_{n+2}), 
	\end{align}
where $X_n = \langle S_n^x \rangle / S$ and $Y_n = \langle S_n^y \rangle / S$. The ground state phase diagram is obtained by minimizing the energy with respect to $X_n$ and $Y_n$. 

For small $\tilde{g}$, the system is in the paramagnetic phase where the spins are polarized along the $z$ axis by the external field term. For $\tilde{g}>\tilde{g}_c$, the system enters distinct magnetic phases. Fig. \ref{spin8}(a) shows the phase diagram in the $\theta-J_1/J_2$ plane for $N=8$. Each magnetic phase has an equivalent superradiant phase in the quantum Rabi chain. The distinct magnetic phases can be described as follows.

(\textit{i}) \textit{Ferromagnetic phase (FP)}. For strong ferromagnetic NN coupling ($J_1 < 0$) with $|\theta_{c,1}^{-}| \leq |\theta| \leq \pi $, the $XY$-Heisenberg interaction favors spin polarization along either the $x$ or the $-x$ axis, yielding identical $X_n$ with $Y_n=0$ in Fig.~\ref{spin8}(b). The system forms a FP with the ferromagnetic ordering, which corresponds to the FSR phase of the quantum Rabi chain.

(\textit{ii}) \textit{Antiferromagnetic phase (AFP)}. For strong antiferromagnetic NN coupling ($J_1 > 0$) with $|\theta_{c,1}^{+}| \leq |\theta| \leq \pi $, the $XY$-Heisenberg interaction favors antiparallel alignment of NN spins along the $J_1$ direction. It results in $X_n=-X_{n\pm1}$ and $Y_n=0$, which is analogous to a real $\alpha_n$ with $\alpha_n=-\alpha_{n\pm1}$ in the AFSR phase of the quantum Rabi chain. Thus, the system enters the AFP with antiferromagnetic ordering between NN sites. Fig.~\ref{spin8}(b) shows the spin configuration $(X_n,Y_n)$, where the spins on NN sites are antialigned.

(\textit{iii})\textit{Chiral magnetic phase (CP)}. For $|\theta_{c,2}^{-}| \leq |\theta| \leq |\theta_{c,1}^{-}| $ and $J_1<J_{1,c}^-<0$, the DM interactions between the NNN cavities dominate over the $XY$ coupling. The spins become noncollinearly polarized with nonzero $X_n$ and $Y_n$, which corresponds to complex $\alpha_n$ of the CSR phase in the quantum Rabi chain. The resulting in-plane magnetization  orientation in the $xy$ plane is shown in Fig.~\ref{spin8} (b), signaling the CP. For $J_1 < 0$, the odd and even sublattices are connected ferromagnetically, and the configurations of site pairs ($1,2$), ($3,4$), ($5,6$), and ($7,8$) exhibit similar orientations. 

(\textit{iv}) \textit{Antichiral magnetic phase (ACP)}. For $|\theta_{c,2}^{+}| \leq |\theta| \leq |\theta_{c,1}^{+}| $ and $J_1>J_{1,c}^+>0$, the system exhibits noncollinear spin polarization with nonzero $(X_n,Y_n)$. Unlike the CP phase, the antiferromagnetic NN coupling ($J_1 > 0$) favors antiparallel alignment between NN sites, which is similar to the configuration of complex $\alpha_n$ in the ACSR phase of the quantum Rabi chain. Fig.~\ref{spin8}(b) shows the in-plane magnetization in the $xy$ plane, exhibiting opposite orientations on the odd and even sublattices. This phase is referred to as the ACP.

(\textit{v}) \textit{Stripe antiferromagnetic phase (sAFP)}.  For $0 \leq |\theta| \leq |\theta_{c,2}^{\pm}| $ and $J_{1,c}^-<J_1<J_{1,c}^+$, the $XY$ coupling between NNN sites dominates over that of NN sites. Spins along the $J_2$ direction become antiparallel, while spins in the $J_1$ direction are parallel. Fig.~\ref{spin8}(b) shows that the configurations of site pairs ($1,2$), ($3,4$), ($5,6$), and ($7,8$) are parallel, in contrast to the conventional AFP phase. This phase is referred to as the sAFP phase.
    
Fig.~\ref{spin8}(c) shows the magnetic phase diagram for $N=6$. Compared with the $N=8$ case, the sAFP phase vanishes as a result of sublattice geometric frustration. Moreover, the CP and ACP phases feature distinct noncollinear spin configurations satisfying $Y_1=Y_4=0$, as shown in Fig.~\ref{spin8}(d). This is reminiscent of the $B_1=B_4=0$ condition found in the quantum Rabi chain. 
    
We thus obtain analogous phase diagrams for the quantum optical system and the effective magnetic system. Each magnetic phase, characterized by the values of $X_n$ and $Y_n$, corresponds to an equivalent phase in the quantum Rabi zigzag chain, described by $A_n$ and $B_n$. This similarity not only implies that the quantum Rabi chain can simulate magnetic behavior but also enables a deeper understanding of chiral optical phases through comparison with chiral magnetic phenomena.
 
\section{Conclusion}
We investigate the quantum phases of the quantum Rabi zigzag chain subjected to an artificial magnetic field, which is realized by modulating the phase of the NNN photon hopping amplitude. We present an analytic solution for the phase boundary in the phase diagram. We uncover chiral and antichiral superradiant phases with opposite photon currents in the ground state, which are distinguished from the  FSR and AFSR phases with no current. We find that geometric frustration within each triangular plaquette favors the stabilization of the chiral phase even at small magnetic flux. When the NNN coupling dominates the NN interaction in the small-flux regime, a sAFSR phase emerges with antiparallel NNN displacements, unlike the conventional AFSR phase. In the $N=6$ system, this stripe phase is suppressed by geometric frustration within the sublattices. 

Remarkably, we map the quantum optical Hamiltonian onto an effective magnetic Hamiltonian containing $XY$-Heisenberg and DM interactions. In particular, the DM interaction, which originates from the artificial magnetic field, is responsible for the chiral phase and noncollinear spin textures. The effective magnetic model displays rich magnetic phases that are analogous to the anomalous superradiant phases in the quantum optical system. The order parameters in these superradiant phases can be interpreted in terms of familiar spin configurations. The mapping between the two systems enables the simulation of chiral magnetism in a highly tunable quantum optical setting.

\section{Acknowledgments}
This work was supported by the National Natural Science Foundation of China under Grants No. 12475013, No. 12474140, No. 12404145, and No. 12547101, the Fundamental Research Funds for the Central Universities (No. 2025CDJ-IAISYB-029), and the China Postdoctoral Science Foundation (No. 2025M783465).

\bibliographystyle{iopart-num}
\bibliography{refs}

\end{document}